# Dynamic patterning and texture evolution of microbubbles in non-Newtonian immobile droplet

Mayuri Bora, Sushant K. Behera and Pritam Deb*

*Advanced Functional Material Laboratory (AFML), Department of Physics, Tezpur University (Central University), Tezpur-784028, India.*

*Corresponding author

Email address: pdeb@tezu.ernet.in (Pritam Deb)

The dynamical perspectives of bubble in a liquid droplet on smooth solid substrate can be revealed by investigating interfacial self-assembly phenomena. Moreover, the complexity in such system can be scaled down into a transparent immobile non-Newtonian droplet. Here we present the dynamical approaches of internal microbubbles through coalescence and implosion near the edge of the droplet. The aggregation density of microbubbles near the rim is greater than the central region corresponding to the dynamical behaviour of the droplet marked with capillary flow along the triple phase contact line. We have also numerically analysed the occurrence of aggregation density of internal microbubbles near the edge by implementing boundary integral method. Our numerical results show good agreement with the experimental findings until the aggregation density have occurred near the edge of the droplet. This understanding ascribes distinctive skew exponential power law characteristics through coalescence and implosion.

Dynamical behaviour of bubbles is a naturally occurring phenomenon and internal microbubbles self-assemble within a transparent non-Newtonian immobile droplet following this dynamical behaviour. The effectiveness of bubble dynamics can play an important role in revealing the bubble size distribution. The phenomenon of self-assembled bubble dynamics has important application in engineering and technological aspect which includes microfluidics[1], foam formation[2], froth floatation[3], emulsion[4]. The dynamic evolution of bubbles has been investigated in various polymers[5,6], surfactant solution[7,8], bubble column reactors[9], salts[10,11] and solvents.[12,13] In spite of experimental[14] and theoretical predictions[15] on bubble self-organization, a very little attention was paid to understand spatial and temporal behaviour of the bubble dynamics in a transparent immobile droplet where the self-assembly of microbubbles are associated near the edge. In recent times, the bubble dynamics investigations have been realized in bubble column reactors[16], population balance framework[17], x-ray microscopy[18], foams.[19,20] In broad classification, the bubble dynamics has been explored in Newtonian[21] and few perspectives in non-Newtonian fluids.[22] Here, evolution of bubbles has been investigated in a non-Newtonian outer fluid and the volume of Fluid-Continuum Surface Force (VOF-CSF) numerical method[23] was employed to understand the coalescence and fragmentation process of three or four bubbles. In another attempt, the coalescence and fragmentation processes were realized under the bubble column reactor where a syringe pump was attached to create gas bubbles manually.[24] The coalescence preference study has been understood by implementing the power-law relationship of a coalesced sphere related to parent size ratio of both bubbles and droplet.[25] In these studies, the bubble dynamics and related phenomena have been explored by incorporating the external effect (injection process, generation of gas bubbles in electrolytes[26] etc.) for the formation of bubbles and droplets in a chamber.

Self-assembly, followed by dynamical behaviour, is a common and ubiquitous phenomenon in bubbles[27,28] and droplets.[29] When two spherical bubbles come in contact, a larger bubble forms which results in minimization of the surface area. The hydrodynamics of bubbles was explained by mass transport phenomena.[30,31] Though few prospects of non-Newtonian fluid have been studied[22,23], but the same fluid has not been incorporated in understanding the bubble dynamics in a static droplet. Studies were made to understand the immobile droplet in dynamical behaviour of drying phenomena. Droplet evaporation and drying phenomena were hypothesized through various mechanisms and models.[32,33] In contrast to these studies, the present manuscript sheds light on other aspect of the drying phenomena i.e. the hydrodynamics and mass transport phenomenon, which includes self-assembled spatiotemporal nature of internal microbubbles in a non-Newtonian immobile droplet. Besides, droplet drying mediated self-assembly mechanism has been rationalized, but microscopic realization and quantitative measurement of internal bubble dynamics are yet to be realized in a comprehensive way. We have adopted a systematic study to reveal the morphological perspectives and texture evolution of the microbubbles near the interface using light microscopy in this current work. Numerical analysis is implemented based on the aggregation density which is incorporated using boundary integral method (BIM). We are focussing on the internal dynamics of microbubbles where the power law fails to describe the distinctive skew exponential characteristics i.e. the diameter of internal bubbles with respect to time expands upto a certain time period and then shrinks leading implosion.

Here, we have conducted the experiment to observe the dynamic nature of the transparent droplet via light microscope. The droplet comprises of multicomponent inorganic materials ($MnCl_2.4H_2O$, NaHSe solution, 3-Mercaptopropionic Acid) organized in water as a solvent. A Light microscope (LEICA DMI 6000 B) with a motorized X-Y stage is employed to observe the dynamic behaviour of an isolated droplet. 0.1 mL of droplet was pipetted from vial and placed in a clean glass slide or cover slip and the experiment was performed in an ambient condition. The glass slide was placed in a motorized X-Y stage. After the droplet was placed, gradually the formation and aggregation of bubbles was occurring near the edge of the droplet which was visualized with the aid of Light Microscope.

Multiple and accurate techniques have been explored to understand the droplet drying phenomena which includes fluorescent microscopy[34], *in situ* optical microscopy[35], bright field optical microscopy[36] for observing the evolution and dynamic nature of the droplet. In this regard, dynamical behaviour of internal bubbles in single transparent immobile droplet has been visualized under the bright field laser light illumination of the light microscope as shown in Fig. 1. Here, a transparent droplet has been taken to study the aggregation occurrence and its subsequent behavioural aspects. Continuous laser light illumination causes the temperature gradient and creates change in thermal environment of the droplet under the focus area with laser source. It is observed that the aggregation density near the contact line with substrate is more concentrated compared to the core region of the droplet. Both thermal perturbation and non-uniformity distribution of aggregation density make non-uniformity in the drying procedure along the droplet interface due to which the aggregation growth near the boundary is showing a random behaviour. Besides, this continuous laser light illumination induces temperature gradient giving rise to the surface tension gradient on the droplet. The gradient of surface tension induces mass-transfer effect at the interface of triple phase contact line (i.e. air-liquid-solid triple phase contact line), where the bubbles inside the droplet follow a random aggregation near the edge in a capillary flow motion. As a result, new bubble formation, coalescence among the existing bubbles and implosion of randomly oriented bubbles take place simultaneously with the drying phenomenon. As a consequence, two bubbles come closer to coalesce and form a bigger bubble and gradually the bubble gets imploded. Besides, detailed illustration of the aggregation density followed by coalescence is shown in (Fig. S1, ESI†) and implosion mechanism is shown in (Fig. S2, ESI†) of the microbubbles inside the immobile droplet and the dynamic behaviour of the microbubbles is captured and provided in Supplementary Information video.

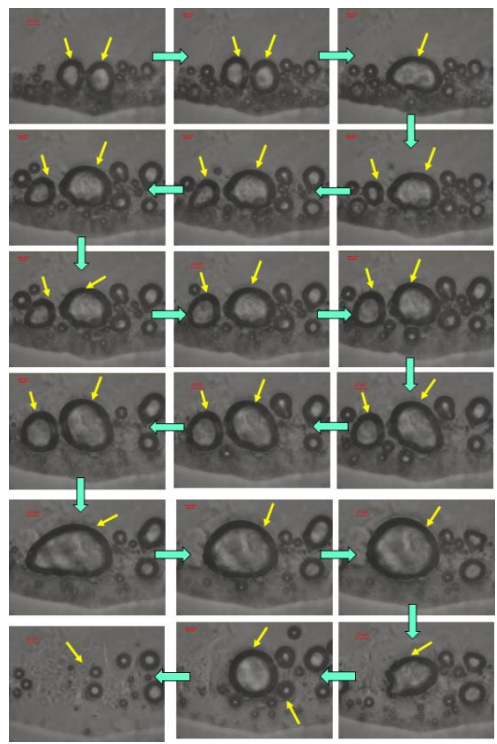

**Fig. 1** Behavioural dynamics of bubbles aggregating near the boundary of the droplet followed by coalescence mediated implosion with time variation.

Fig. 2 represents schematically the mechanism, discussed in Fig. 1, of this dynamical behaviour of the bubbles inside an immobile droplet. The scheme of the dynamic process has been shown in the form of a trapezium covering all the sequential aspects of the internal bubbles entirely from initial stage to implosion stage. Outside the trapezium, it depicts the coalescence and implosion of bubble with intermediate stage which is visualized with the aid of light microscope and the highlighted blue oval shape (at the bottom portion of the trapezium) depicts the stages of droplet.

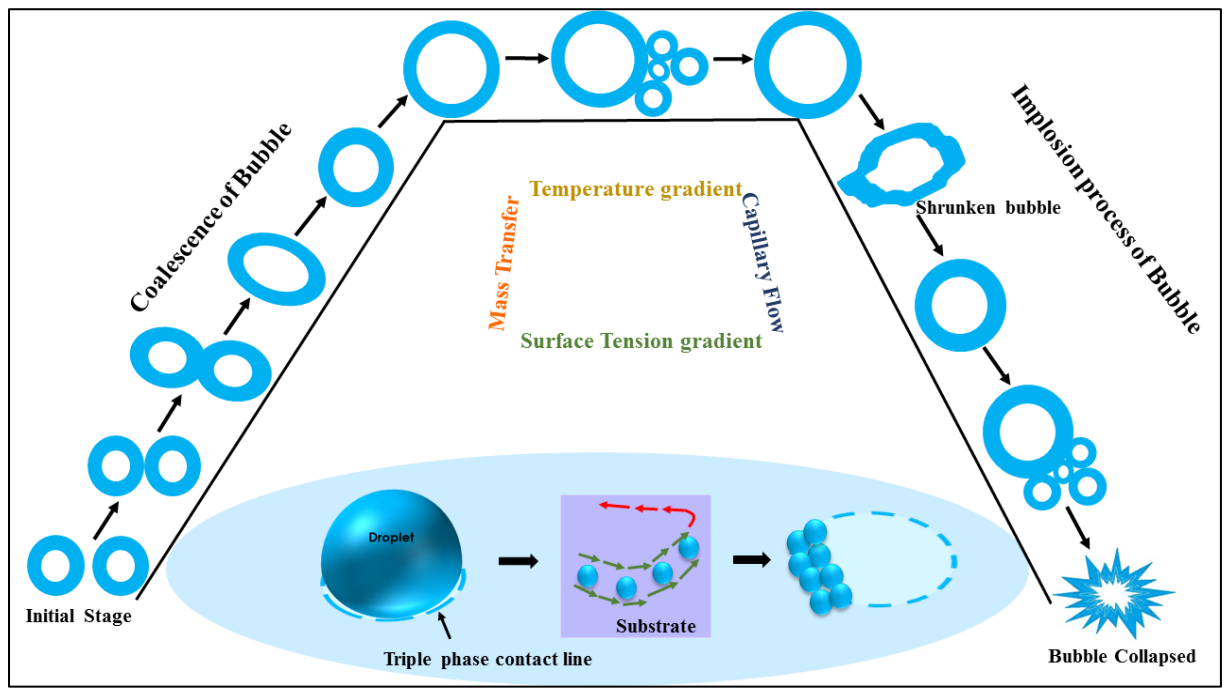

**Fig. 2** Schematic representation of the coalescence and implosion process of internal microbubbles inside the immobile droplet.

In this experiment, the droplet has been placed on a glass surface under ambient condition. As discussed earlier, the surface tension gradient results non-uniform mass transfer during evaporation of the droplet with higher value of transfer from the boundary region compared to the core region of the same droplet. Thus, temperature and surface tension gradients are realized by implementing mass-transfer effect, between the core and boundary of the droplet induced at the air-liquid-glass (triple) phase contact line. Therefore, it tunes the flow of internal bubbles between the edge and the core of the droplet. Besides, it is noticed that the random behaviour is occurring in a capillary motion towards the core of the droplet. As a result, the coalescence and implosion of bubble take place simultaneously due to random capillary flow with respect to time which is depicted outside the trapezium.

Dynamical behaviour of the transparent droplet is also evidenced in quantitative estimation of diameter of the internal bubbles with respect to the time as shown in (Fig. 3 (a)). Concurrent stages of coalescence and implosion mechanism of internal bubbles are presented as the inset of (Fig. 3(a)). In the curve, the dotted line represents the quantitative diameter of the bubble with respect to time and the red curve is the fitted curve which is expressed as follow-

$$D = D_0 + Ce^{-0.016t^2} \qquad (1)$$

where, $D_0$ is the initial diameter of the bubbles and C is the amplitude of the curve. In initial stage of 10 sec, the two bubbles come closer gradually and it gets coalesce to form an oblate shape. Meanwhile oblate shape is not stable for longer time considering its surface tension value, unlike the spherical shape. Thus, spherical shape is taken by the bubble without much shape variation, but the thickness slightly increases to form a bubble with maximum diameter at 134 sec, afterwards, the diameter contracts gradually due to the simultaneous evaporation process. At around 170 sec, the bubble undergoes to take the smallest diameter where it again takes a spherical shape due to surface tension and the coalescence of smaller bubble. Finally the spherical bubble has been imploded at around 250 sec.

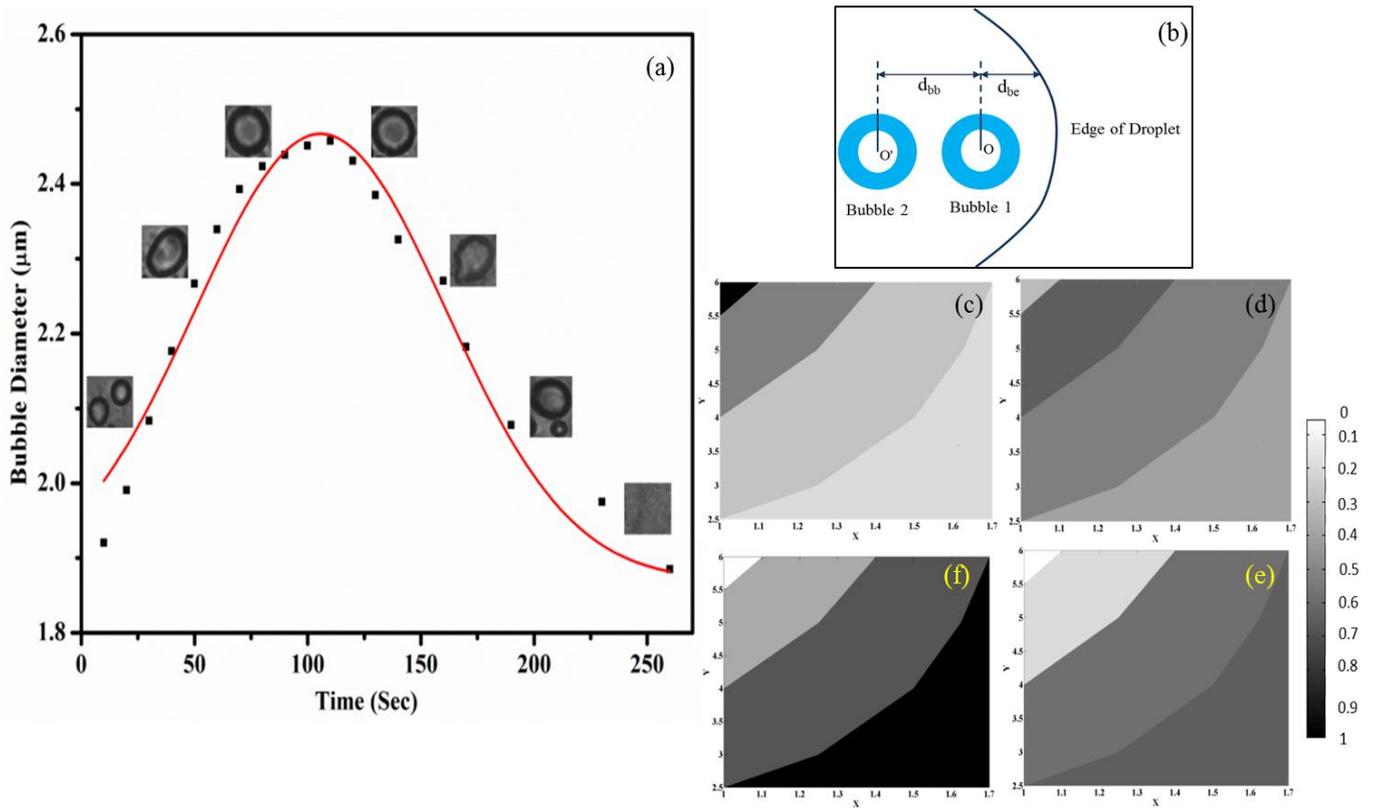

**Fig. 3** (a) Skew-exponential power law curve representing the variation in diameter of microbubbles as a function of time. (b) Numerical Model of a droplet where two bubbles are aggregated near the boundary. (c, d, e, f) aggregation density profile of internal microbubbles from core to the edge and their evolution with respect to time.

The BIM is implemented to understand the coalescence and implosion of internal bubbles inside the droplet (shown in the model Fig. 3(b)) following the numerical analysis technique of two non-buoyant bubbles[37]. In the model, the bubble nearer to the interface is denoted as Bubble 1, while the bubble next to it is denoted as Bubble 2. The distance between the edge of the droplet and initial centre of bubble 1 is $d_{be}$, the initial inter-bubble distance is denoted by $d_{bb}$, and the maximum radius of bubbles (i.e., $[3V_{max}/(4\pi)]^{1/3}$, where $V_{max}$ represents the maximum volume of bubble) are denoted by $R_{m1}$ and $R_{m2}$. In our report, we are primarily concerned with the coalescence and implosion of bubbles near the edge of the droplet. The shape of the bubble is given by;

$$\frac{dx}{dt} = u(x,t), x \, \varepsilon \, S, \qquad (2)$$

where, **x** is a position vector, S the bubble surface, **u** the velocity vector and t is the time. Here, we consider the droplet to be an incompressible and inviscid fluid where the velocity is interconnected to the pressure gradient($\nabla \varphi$) by;

$$u = u_0 - \frac{1}{\rho}\int_{x_0}^{x} \nabla \varphi \, dt, \qquad (3)$$

where, ρ is the density of the droplet.

In this work, we follow a theory to model the aggregation density of internal non-buoyant bubbles near the edge of the droplet monitored by coalescence and implosion of bubbles. As the droplet is considered to be inviscid and incompressible, thus, the flow is governed by the Laplace equation leading to the velocity being expressed as the gradient of pressure φ, i.e. $u = \nabla\varphi, \nabla^2\varphi = 0$. This equation follows the boundary integral equation, $u(x,t).\varphi(x,t) = \iint_S [\frac{\partial \varphi(q)}{\partial n} G(x,q) - \varphi(q)\frac{\partial}{\partial n} G(x,q)]ds$, where, **x** is the position vector, q is the integral point on the boundary of the droplet and $\frac{\partial}{\partial n}$ is the normal outward derivative from the boundary of the droplet and the Green function G(**x**,**q**) is taken as $G(x,q) = \frac{1}{|x-q|} + \frac{1}{|x-q'|}$ where, **q'** is the reflected image of **q** across the boundary of the droplet. The dynamic boundary condition on the surfaces of bubbles is in dimensionless forms, $\nabla^2\varphi = 0$, for a>0, b>0 and $a^2 + b^2 < 1$ which is subjected to the boundary condition, $\varphi = b$ on $a = 0$ for 0<b<1.

$$\varphi = a + b \text{ on } a^2 + b^2 = 1, \text{ for } a > 0, b > 0 \qquad (4)$$

$$q = \frac{\partial \varphi}{\partial n} = -1 \text{ on b = 0 for 0<x<1}$$

The Boundary Integral equations have been derived from the Laplace's equation considering the boundary conditions as shown in eqn. (4).

Fig. 3(b) depicts the model of two bubbles near the edge of the droplet. Fig. 3(c) depicts the core of the droplet is more dense with the aggregates which is shown with a black colour and the other part of the droplet is shown with the lighter colour explaining the less dense area of the droplet. Moreover, it is clear that the aggregation of internal bubbles is gradually tending towards the boundary with the change in the colour gradient where the core of the droplet is gradually occurring less dense than the boundary as shown in Fig. 3(d) and 3(e)). Fig. 3(f) shows that the core of the droplet have become less dense whereas the aggregation density is higher near the boundary of the droplet.

In summary, this work reports the comprehensive results elucidating the bubble dynamics of a transparent non-Newtonian immobile droplet. The mechanism behind the self-assembly of bubbles has been associated with temperature gradient and mass transfer along the interface which arises due to surface tension gradient. The aggregation density of bubbles near the rim is dominating over the density at the central region conforming the dynamic behaviour of the internal microbubbles along the triple phase contact line. The accumulation of bubbles near the edge leads to dynamic coalescence and implosion process. The dynamic behaviour shows an expansion of bubbles at 134 s, contraction at 170 s and disappears at 250 s which followed the coalescence and implosion mechanism of microbubbles inside the droplet. Using Boundary Integral Method based numerical technique, the aggregation density profiles of internal bubbles are shown with the help of colour gradient at different stages of aggregation depicted near the interface of the droplet. Here, unlike the power-law characteristic, the skew-exponential power law, which describes different stages of coalescence and implosion of internal microbubble with time, is predominant.

The authors acknowledge to the UGC research award from University Grants Commission, Govt. of India, vide grant no. F.30-1/2014/RA-2014-16-GE-WES-5629 (SA-II). MB acknowledges Tezpur University for providing financial support.

**Conflicts of interest**

There are no conflicts to declare.

**Notes and references:**